\documentclass[sigconf,nonacm]{acmart}

\usepackage{algorithmic}
\usepackage{graphicx}
\usepackage[skip=5pt]{caption}
\usepackage{textcomp}
\usepackage{xcolor}
\usepackage{enumitem}

\renewcommand{\baselinestretch}{0.99}

\newcommand{\vscc}{vscc}
\newcommand{\mvcc}{mvcc}
\newcommand{\bmac}{BMac}
\newcommand{\protocolprocessor}{protocol\_processor}
\newcommand{\bmp}{\bmac{} protocol}
\newcommand{\dataremover}{DataRemover}
\newcommand{\annotator}{AnnotationGenerator}
\newcommand{\packetprocessor}{PacketProcessor}
\newcommand{\datainserter}{DataInserter}
\newcommand{\dataextractor}{DataExtractor}
\newcommand{\dataprocessor}{DataProcessor}
\newcommand{\hashcalculator}{HashCalculator}
\newcommand{\datawriter}{DataWriter}

\newcommand{\blkprocessor}{block\_processor}
\newcommand{\blkmonitor}{block\_monitor}
\newcommand{\regmap}{reg\_map}
\newcommand{\blkfifo}{block\_fifo}
\newcommand{\txfifo}{tx\_fifo}
\newcommand{\Endsfifo}{ends\_fifo}  
\newcommand{\rdsetfifo}{rdset\_fifo}
\newcommand{\wrsetfifo}{wrset\_fifo}
\newcommand{\resfifo}{res\_fifo}
\newcommand{\ecdsaeng}{ecdsa\_engine}
\newcommand{\blkverify}{block\_verify}
\newcommand{\blkvalidate}{block\_validate}
\newcommand{\txscheduler}{tx\_scheduler}
\newcommand{\txverify}{tx\_verify}
\newcommand{\txvscc}{tx\_vscc}
\newcommand{\txvalidators}{tx\_validators}
\newcommand{\Endsscheduler}{ends\_scheduler}
\newcommand{\Endspolicyeval}{ends\_policy\_evaluator}
\newcommand{\txcollector}{tx\_collector}
\newcommand{\txmvcc}{tx\_mvcc\_commit}
\newcommand{\swvalidator}{sw\_validator}

\begin{document}
\title{Blockchain Machine: A Network-Attached Hardware Accelerator for Hyperledger Fabric}

\author{Haris Javaid}
\email{harisj@xilinx.com}
\affiliation{
	\institution{Xilinx, Singapore}
	\country{}
}
\author{Ji Yang}
\email{jamiey@xilinx.com}
\affiliation{
	\institution{Xilinx, USA}
	\country{}
}
\author{Nathania Santoso}
\email{nathania@xilinx.com}
\affiliation{
	\institution{Xilinx, Singapore}
	\country{}
}
\author{Mohit Upadhyay}
\authornote{Work done during internship at Xilinx. Currently at National University of Singapore, email: mohitu@u.nus.edu.}
\email{mohitu@xilinx.com}
\affiliation{
	\institution{Xilinx, Singapore}
	\country{}
}
\author{Sundararajarao Mohan}
\email{mohan@xilinx.com}
\affiliation{
	\institution{Xilinx, USA}
	\country{}
}
\author{Chengchen Hu}
\email{chengche@xilinx.com}
\affiliation{
	\institution{Xilinx, Singapore}
	\country{}
}
\author{Gordon Brebner}
\email{gjb@xilinx.com}
\affiliation{
	\institution{Xilinx, USA}
	\country{}
}

\begin{abstract}
In this paper, we demonstrate how Hyperledger Fabric, one of the most popular permissioned blockchains, can benefit from network-attached acceleration. The scalability and peak performance of Fabric is primarily limited by the bottlenecks present in its block validation/commit phase. We propose Blockchain Machine, a hardware accelerator coupled with a hardware-friendly communication protocol, to act as the validator peer. It can be adapted to applications and their smart contracts, and is targeted for a server with network-attached FPGA acceleration card. The Blockchain Machine retrieves blocks and their transactions in hardware directly from the network interface, which are then validated through a configurable and efficient block-level and transaction-level pipeline. The validation results are then transferred to the host CPU where non-bottleneck operations are executed. From our implementation integrated with Fabric v1.4 LTS, we observed up to 12$\times$ speedup in block validation when compared to software-only validator peer, with commit throughput of up to 68,900 tps. Our work provides an acceleration platform that will foster further research on hardware acceleration of permissioned blockchains.
\end{abstract}
\maketitle

\section{Introduction}\label{sec:introduction}
As enterprises adopt blockchain technology, permissioned block\-chains are becoming the de facto blockchain platform for enterprise applications. In permissioned blockchains, unlike public blockchains such as Bitcoin and Ethereum, only known nodes participate in the blockchain network (because the identity of nodes is authenticated cryptographically) and the consensus is delegated to a few select nodes. These characteristics coupled with builtin data integrity (due to immutable transactions) and data replication (due to distributed ledger maintained by each node) make permissioned blockchains very suitable for enterprise applications. Many permissioned blockchains are available such as Hyperledger Fabric~\cite{HyperledgerFabric}, Quorum~\cite{Quorum} and Corda~\cite{Corda}. Fabric is one of the most popular permissioned blockch\-ains as it is open-source and has already been used to implement and deploy many production-ready enterprise applications in finance and supply chain domains~\cite{Castillo2021,Ehrlich2021}.

\textbf{Motivation.} One of the major challenges facing permissioned blockchains is their scalability to achieve performance required for real-life applications. For example, Fabric has been shown to deliver throughput from a few hundred to thousands of transactions per second (tps): ${\sim}$600 tps for create\_asset benchmark~\cite{Hyperledger2020FabricPerformanceReports}, and ${\sim}$7,700 tps~\cite{Thakkar2021} and ${\sim}$20,000 tps~\cite{Gorenflo2019FastFabric} for smallbank benchmark~\cite{HyperledgerCaliperBenchmarks}. Although impressive compared to Bitcoin's ${\sim}$5 tps~\cite{Blockchair2020BitcoinThroughput} and Ethereum's ${\sim}$15 tps~\cite{Blockchair2020EthereumThroughput}, Fabric's throughput falls short of the required performance of real-life applications such as payments through Visa, which must be capable of processing more than 65,000 tps under peak workload~\cite{Visa2017}. This calls for further performance improvements to aid in scalability of permissioned blockchains and their more widespread adoption, which has been pointed out in recent literature~\cite{Thakkar2021} and is the focus of this paper.

Hyperledger Fabric uses the \textit{execute-order-validate} model, where a transaction is executed/endorsed first, then ordered into a block, which is finally validated/committed to the ledger. Consequently, some nodes in the Fabric network act as endorser peer to execute/endorse transactions and validate/commit blocks or validator peer (also known as non-endorsing peer) to only validate/commit blocks, while other nodes act as orderers to create new blocks. It is well-known that validation phase (i.e., validator peer, and not the consensus mechanism which creates blocks) is one of the major bottlenecks and critically affects the achievable peak throughput~\cite{Gorenflo2019FastFabric,Javaid2019,Chung2019,Thakkar2021}. These works also proposed performance improvements for validation phase, such as parallel/pipelined validation of transactions, caching block data, partial validation/commit of blocks, and separating endorsement and validation phases. However, all these optimizations are purely software-based, and hence the resulting performance improvements are limited by the computational power of underlying general-purpose compute resources such as multi-core servers.

\textbf{Goals.} In this paper, we improve performance of Hyperledger Fabric's validator peer beyond what is currently achievable with software implementations, by leveraging hardware-based acceleration using FPGA accelerator cards. Fabric is available as Blockchain-as-a-Service from all major cloud providers (e.g., IBM Blockchain Platform, Amazon Managed Blockchain, etc.), and is one of the top blockchains~\cite{Gartner2021} with hundreds of enterprise applications already developed and deployed. FPGA accelerator cards are being increasingly adopted for accelerating cloud workloads~\cite{Xilinx2020DataCenter} and are also available from major cloud providers such as AWS and Microsoft Azure. Therefore, we accelerate core operations of Fabric using FPGAs, and our work can benefit all Fabric based applications.

Since Fabric is a very complex distributed system with many components, designing a hardware accelerator requires careful consideration of the following challenging goals:
\begin{itemize}[leftmargin=2.5ex]
	\item \textbf{High performance:} The validation phase involves various types of operations such as verification of cryptographic signatures, unmarshaling of protocol buffers~\cite{Varda2008}, reading from/writing to database, and writing to disk-based ledger. To achieve high throughput, it is important to analyze the validation phase in detail to find out operations most suitable for hardware acceleration, while keeping non-critical operations in software. Network-attached accelerators (where data goes directly from the network interface to accelerator card) significantly outperform traditional CPU-centric architectures (where data goes from the network interface to accelerator card via host CPU)~\cite{Weerasinghe2016,Tork2020}. In Fabric and other blockchains, block data comes from the network, hence network-attached accelerators are a natural choice. A communication mechanism (bypassing the host CPU) for efficient transfer of block data and its access by hardware accelerator is also needed. Although high throughput is the primary goal, the following goals are essential for practical deployment/adoption.
	\item \textbf{Adaptability:} Fabric uses smart contracts (also known as chaincodes) for implementation of business logic, which means that many diverse applications can be implemented. The hardware accelerator should be configurable based on the applications that would be deployed. For example, a banking application may have a very different cryptographic workload (a function of chaincode endorsement policy~\cite{Thakkar2018}) than a supply chain application. FPGAs provide the flexibility for custom parallelization and pipelining of critical operations, and custom data movement and streaming networking, resulting in a much more optimized architecture compared to using fixed-configuration GPUs. Furthermore, programmability of FPGAs can create application-specific accelerators with lower cost and effort compared to ASICs. Therefore, we focus on an FPGA based accelerator in this work. Note that ASICs, GPUs and FPGAs are typically used for mining in public blockchains rather than permissioned blockchains.
	\item \textbf{Compatibility:} Each Fabric node acts either as an endorser peer or validator peer or an orderer, which means that any validator peer with hardware accelerator must be compatible with the software-only endorser peers and orderers in order for the Fabric network to work. This is also useful when all the validator peers of an already deployed Fabric network cannot be upgraded with hardware accelerator simultaneously, hence calling for upgrade over multiple cycles (i.e., some validator peers are software-only while others are upgraded with hardware accelerator).
\end{itemize}

\textbf{Contributions.} We propose Blockchain Machine (\textbf{\bmac{}}), a hardware/software co-designed platform with hardware accelerator and modified Fabric software to act as the hardware-accelerated validator peer in a Fabric network (see Figure~\ref{fig:blockchain_machine}). The \bmac{} peer is targeted for a server with network-attached FPGA card (such as Xilinx Alveo~\cite{Xilinx2020Alveo}) in contrast to existing nodes which run Fabric software on just a multi-core server. The \bmac{} peer receives blocks from the orderer through a hardware-friendly protocol, and the block data is retrieved in hardware directly from the network interface. The extracted block and its transactions are then passed through an efficient block-level and transaction-level pipeline, which implements the bottleneck operations of the validation phase. Finally, Fabric software accesses the block validation results from hardware, and then commits the block to ledger just like the software-only validator peer. In particular, our contributions are:
\begin{itemize}[leftmargin=2.5ex]
	\item A \textbf{novel hardware-friendly communication protocol} is proposed (Section~\ref{sec:protocol_processor}), which breaks the block into multiple sections, removes redundancy among the sections, and transmits the data in self-contained UDP packets. Each packet contains annotations in L7 header to specify the type of data to expect in the payload, enabling efficient parsing of a block in hardware directly from the network interface. The \bmac{} peer includes a protocol processor in hardware which reconstructs the block from its sections, and extracts the relevant data for the validation phase operations. Using self-contained UDP packets enables cut-through processing, with a small resource footprint of the protocol processor (compared to TCP protocol where buffering in hardware is needed to receive the entire block before processing it). On the software side, our protocol is implemented in Fabric to send blocks from the orderer to \bmac{} peer, without modifying Fabric's existing Gossip protocol (based on gRPC/HTTP2/TCP). Thus, we use our protocol only when its supported by both sender and receiver (orderer to \bmac{} peer), otherwise we fall back to Gossip protocol (orderer to software-only peers). This makes \bmac{} peer compatible with other Fabric nodes.
	\item A \textbf{novel block processor} is proposed in hardware (Section~\ref{sec:block_processor}), which consists of an integrated block-level and transaction-level pipeline to validate a block and its transactions, and commit valid transactions to the database. The transaction-level pipeline uses parallel-pipelined architecture to validate multiple transactions in parallel as well as pipelined fashion to achieve high throughput. The block processor implements the most critical operations of the validation phase~\cite{Thakkar2021,Javaid2019}: verification of cryptographic signatures, verification of endorsements and endorsement policy, and reading from/writing to database.
	\item A \textbf{hardware/software co-designed platform} (Section~\ref{sec:overview} and Section~\ref{sec:hwsw_interface}), where block data streams from the network interface to protocol processor and block processor, coupled with an API which is used by the Fabric software to access the results of block validation. The \bmac{} architecture is adaptable as it can be customized for the applications (e.g., software-defined endorsement policy is compiled into the hardware, or the number of transactions to validate in parallel is configurable).
\end{itemize}
We implemented \bmac{} peer on a server with network-attached Xilinx Alveo U250 FPGA card, and integrated it with Hyperledger Fabric v1.4 LTS. Our experiments show that the commit throughput of \bmac{} peer ranges from 9,200 -- 68,900 tps for smallbank and drm benchmarks~\cite{HyperledgerCaliperBenchmarks}, compared to 2,500 -- 7,200 tps of software-only validator peer. We plan to open-source our \bmac{} peer and its upgrade to Fabric v2.2 LTS is in progress. We believe that our work provides the necessary platform to enable further research on hardware acceleration of permissioned blockchains.

\section{Permissioned Blockchains}
\begin{figure}[b]
\vspace{-3ex}
	\centering
	\includegraphics[width=0.95\columnwidth]{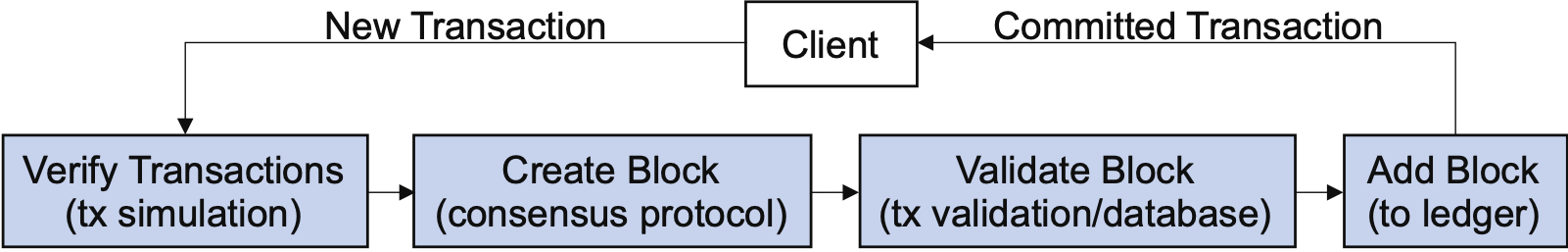}
	\caption{Generalized Flow of a Permissioned Blockchain.}
	\label{fig:permissioned_blockchain}
\end{figure}

A permissioned blockchain is a peer-to-peer network of nodes, where each node executes certain operations.  Figure~\ref{fig:permissioned_blockchain} depicts a generalized flow of how transactions are processed by a permissioned blockchain. A new transaction is created by a client and sent to nodes that are responsible for verification of transactions. These nodes verify that the client is authorized to interact with the network, and simulate the transaction to ensure that its executable and can be committed later. A block is created from verified transactions using a consensus protocol, where by a group of nodes agree upon the order of transactions in a single block (and the order of blocks in the ledger). The newly created block is broadcast to all the nodes, where each block and its transactions are validated (checks on cryptographic signatures, permissions, database accesses, etc.). Afterwards, every node adds the block  (thereby committing the transaction) to its own copy of the ledger, resulting in a decentralized/distributed ledger maintained across the network.

Although some of these operations can be implemented in a slightly different order, the noteworthy point here is that all permissioned blockchains involve broadcasting of blocks and extensive use of cryptopgrahic operations and database accesses. For example, permissioned Ethereum (which is a variant of public Ethereum to address requirements of enterprise applications and is available as Hyperledger Besu~\cite{HyperledgerBesu} or Quorum~\cite{Quorum}) combines simulation of transactions with block creation, while Fabric keeps these operations decoupled (tx simulation in endorser peer and block creation in orderer). However, all of Fabric, Besu and Qourum receive blocks from the network, validate blocks and their transactions, and add blocks to the ledger. Therefore, although we target Fabric, our work accelerates core operations (such as block receiving, transaction verification and validation, database accesses, etc.) and hence can be applied to other permissioned blockchains.

\subsection{Hyperledger Fabric}\label{sec:fabric_arch}
\subsubsection{Overview}
Hyperledger Fabric is an open-source, enterprise-grade implementation of a permissioned blockchain. A Fabric network consists of peers, orderers and clients, where each node has an identity provided by membership service provider. Each peer maintains its own copy of the ledger and current global state of the data in a state database (LevelDB~\cite{Triandana2019GoLevelDB} or CouchDB~\cite{CouchDB}). An endorser peer both executes/endorses transactions and validates/commits blocks to the ledger, while a validator peer only validates/commits blocks to the ledger. Transactions invoke smart contracts or chaincodes, which represent the business logic and are instantiated on the endorser peers. The ordering service consists of one or more orderers, which use a consensus mechanism to establish a total order for the transactions. Current consensus is based on Raft~\cite{Ongaro2014}.

\begin{figure}[t]
	\centering
	\includegraphics[width=0.7\columnwidth]{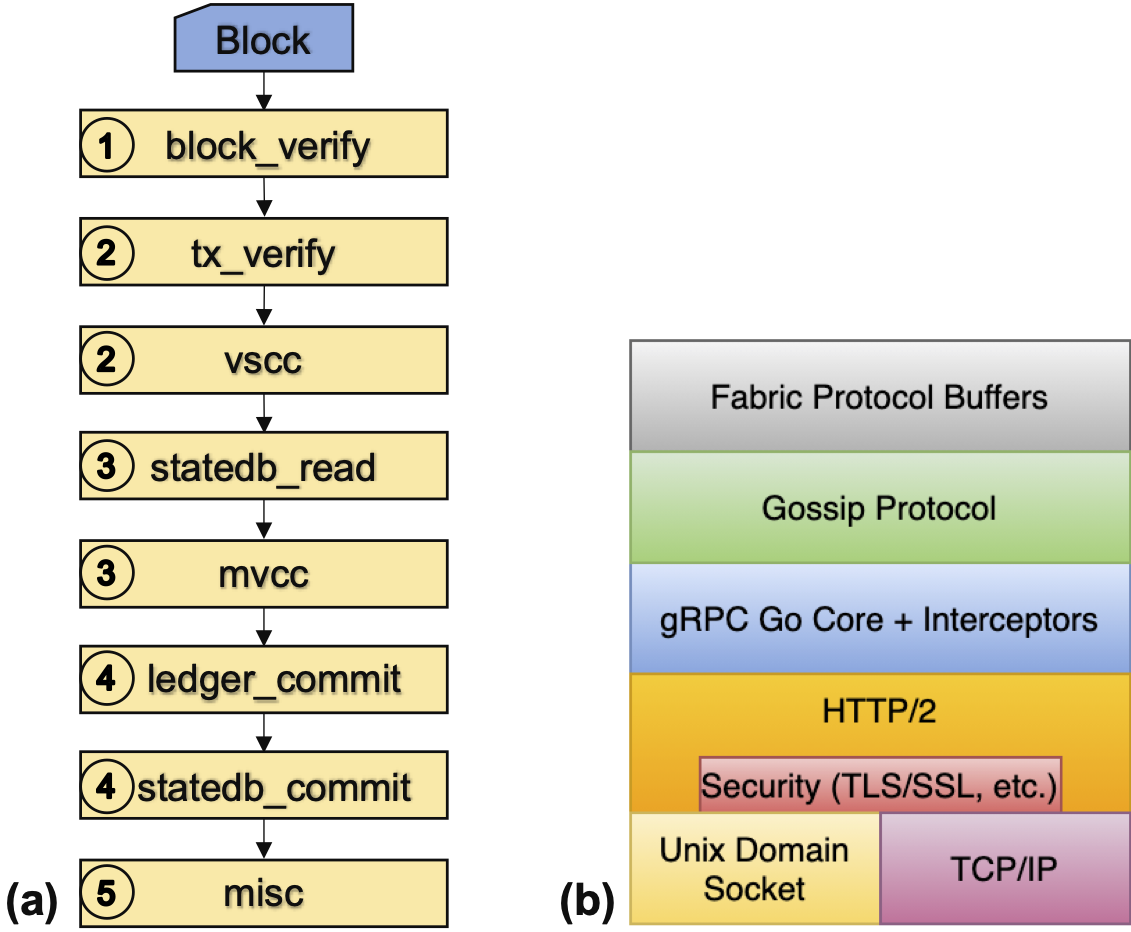}
	\caption{Hyperledger Fabric: (a) Validation Phase (b) Networking Stack.}
	\label{fig:validation_phase_network_stack}
\vspace{-4ex}
\end{figure}

A transaction flows through the Fabric network as follows. A client creates a transaction and sends it to a number of endorser peers {first step in Figure~\ref{fig:permissioned_blockchain}}. Each endorser peer executes the transaction against its own state database, in order to compute the read and write sets. The read set contains the keys accessed and their version numbers, while the write set contains the keys to be updated with their new values. If there are no errors, the peer sends back its endorsement to the client, which consists of the original transaction, its read and write sets, and peer's cryptographic signature. By default, Fabric uses 256-bit ECDSA scheme for signature generation and verification. After the client has gathered enough endorsements, it submits the transaction with its endorsements to the ordering service. A block is created from the ordered transactions, and the orderer signs the block and broadcasts it to all the peers (second step in Figure~\ref{fig:permissioned_blockchain}). Each peer first validates the block and all of its transactions, and then commits it to the ledger and state database (third and fourth steps in Figure~\ref{fig:permissioned_blockchain}).

\subsubsection{Validator Peer}
Figure~\ref{fig:validation_phase_network_stack}a shows validation phase as a pipeline of various operations. After receiving the block from the ordering service (or another peer), in step 1, the peer checks that the block is not a malicious block by retrieving its data and verifying orderer's signature. In step 2, each transaction in the block is verified and validated. The verification of transaction involves retrieving transaction data and verifying client's signature. The validation of transaction involves execution of \vscc{} (validation system chaincode), where the endorsements are verified and the endorsement policy of the associated chaincode is evaluated. An endorsement policy specifies the type and number of endorsers needed for the transaction in the form of logical expressions such as ``Org1 \& Org2'' or ``2-outof-3 orgs'' where org stands for an organization in the Fabric network. A transaction is marked valid only if its endorsement policy is satisfied.

In step 3, \mvcc{} (multi-version concurrency control) check is applied successively to all the valid transactions of the block, starting from the first one. This check ensures that there are no read-write conflicts between the transactions. The read set of each transaction is computed again by accessing the state database, and is compared to the read set from the endorsement phase. If these read sets are different, then the transaction is marked as invalid.

In step 4, the block is finally committed. First, the entire block is written to the ledger with its transactions' valid/invalid flags and a commit hash. Then, the state database is updated with the write sets of all the valid transactions. Step 5 represents miscellaneous operations, such as updating the history database to keep track of which keys have been modified by which blocks and transactions.

Fabric uses a peer-to-peer Gossip protocol for dissemination of data between the orderers and peers. Figure~\ref{fig:validation_phase_network_stack}b depicts Fabric's networking stack. A Gossip message is built using protocol buffers (protobufs), where the block data (including its transactions) is in the form of a marshaled protobuf. The Gossip message is then transmitted through gRPC, which uses HTTP/2 and TCP as its transport layer in Fabric. The validator peer uses the same networking stack to receive the block as a marshaled protobuf, which is then processed through the validation pipeline in Figure~\ref{fig:validation_phase_network_stack}a.

\subsubsection{Bottlenecks in Validator Peer}
\begin{figure}[t]
	\centering
	\includegraphics[width=1.0\columnwidth]{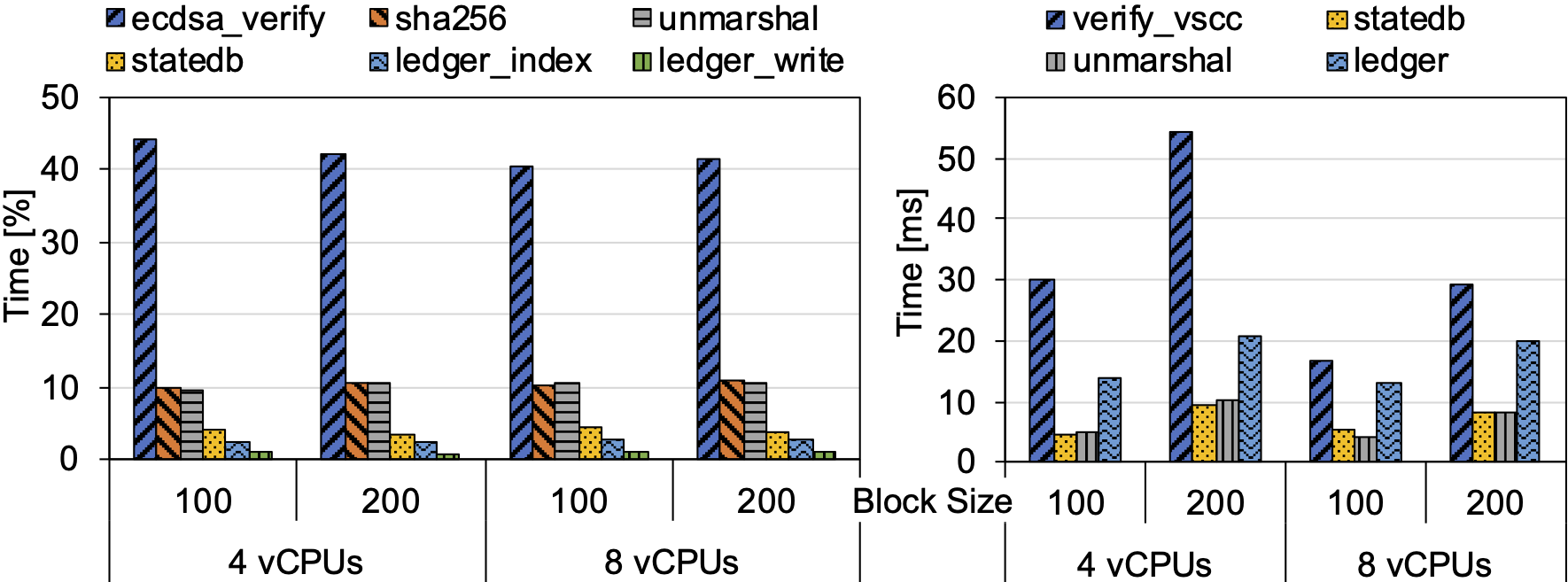}
	\caption{Validator Peer Bottlenecks.}
	\label{fig:bottlenecks}
\vspace{-4ex}
\end{figure}

We conducted extensive experiments to analyze the validation phase, collecting time spent in various operations coupled with in-depth analysis of the call graph. These experiments were conducted under the default setup described in Section~\ref{sec:experiments}. Figures~\ref{fig:bottlenecks}a \& b report the most time consuming operations from profiling (using Go language pprof package) and coarse-grained breakdown of block validation (using timestamp logging around operations in Figure~\ref{fig:validation_phase_network_stack}a) respectively, when block size (number of transactions) and vCPUs (of each validator peer VM) are changed. Our observations are:
\begin{enumerate}[leftmargin=4ex]
	\item Retrieving block and transaction data involves unmarshaling of many protobufs, which is time consuming. Since the validator peer receives the block as a marshaled protobuf, it has to unmarshal the block to access individual transactions for verification, \vscc{} and \mvcc{} operations. From Figure~\ref{fig:bottlenecks}a, the protobuf unmarshaling takes ${\sim}$10\% of the total time, and is the second/third most time consuming operation.
	\item Validation of a block involves verification of many ECDSA signatures, which becomes the critical path. Figure~\ref{fig:bottlenecks}a shows that ecdsa\_verify takes ${\sim}$40\% of the total time, which is much more than any other operation, while ${\sim}$10\% time is spent in computation of SHA-256 hash (of block/transaction data, and is a prerequisite to ECDSA verification). Both these operations significantly contribute to verification and \vscc{} operations in Figure~\ref{fig:bottlenecks}b, where verify\_\vscc{} is always the critical operation. Although increasing vCPUs for the same block size means more transactions can be validated in parallel, resulting in lower verify\_\vscc{} time, the total number of ECDSA and SHA-256 operations remain the same and hence the percentage time in Figure~\ref{fig:bottlenecks}a does not change much.
	\item State database accesses are typically slow, leading to increased validation latency. Although state database accesses are not among the most critical operations, they are still significant: fourth most time consuming operation from Figure~\ref{fig:bottlenecks}a, and takes 10--20\% of the validation time in Figure~\ref{fig:bottlenecks}b. Some state database reads can be done in parallel, while writes are sequential due to \mvcc{} operation, hence adding more vCPUs does not improve state database access time.
	\item Committing to disk-based ledger takes longer for larger block sizes, which could slow down validation. Internally, the ledger commit writes the block to a file and updates the block index (stored in an internal database, and used for checking duplicates). Although ledger commit takes longer than state database access in Figure~\ref{fig:bottlenecks}b, the profiling data in Figure~\ref{fig:bottlenecks}a shows that ledger operations take the least amount of time, which suggests that ledger commit is an I/O bound operation. Furthermore, since blocks are committed sequentially to the ledger, adding more vCPUs does not improve ledger commit time.
\end{enumerate}
These bottlenecks have also been pointed out in various contexts in recent works~\cite{Gorenflo2019FastFabric,Javaid2019,Chung2019,Thakkar2021}, and are the basis for \bmac{} architecture described in the next section.

\section{Blockchain Machine}\label{sec:blockchain_machine}


\subsection{Architecture Overview}\label{sec:overview}
\begin{figure}[b]
\vspace{-2ex}
	\centering
	\includegraphics[width=\columnwidth]{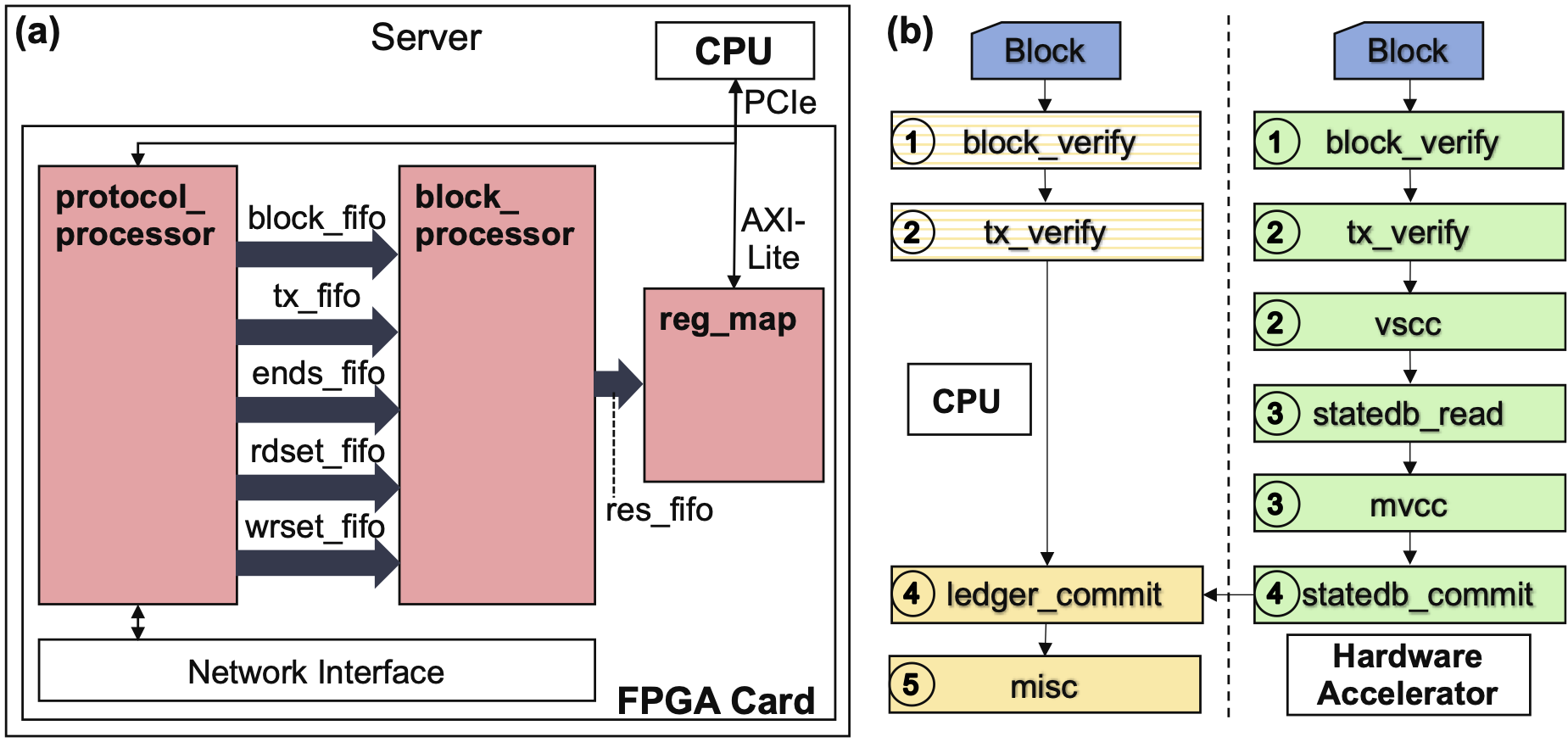}
	\caption{Blockchain Machine: (a) system setup (b) hardware/software partitioned validation phase.}
	\label{fig:blockchain_machine}
\end{figure}

The \bmac{} architecture is based on two principles: (1) efficient transfer and access of block data and its transactions (to address bottleneck 1), and (2) efficient verification and validation of transactions (to address bottlenecks 2 and 3). The ledger commit (bottleneck 4) is an I/O bound operation, hence more suitable for execution on a CPU than a compute accelerator like the \bmac{} architecture; however, ledger commit on CPU is overlapped with the validation operations in hardware for better throughput. Since Fabric's validator peer (and validation nodes in blockchains) receive block data from the network, we designed \bmac{} architecture as a network-attached accelerator (i.e., access and process data bypassing the host CPU, which is better than a traditional CPU-centric architecture~\cite{Weerasinghe2016,Tork2020}). Figure~\ref{fig:blockchain_machine}a illustrates the overall architecture of \bmac{}, in the context of a server with network-attached FPGA card connected to the host CPU through PCIe bus. Although we do not show a Network Interface Card (NIC) attached directly to the CPU, a CPU-attached NIC can be used (to handle non-\bmac{} traffic) as long as the \bmac{} traffic goes through the FPGA card. For the rest of the paper, we assume all the network traffic goes through the FPGA card with integrated network interface, i.e., the FPGA card acts as a NIC.

The \textbf{\protocolprocessor{}} processes all the incoming Ethernet packets, and classifies them as either normal packets or \bmac{} packets (i.e., packets sent through our protocol). All the normal packets are forwarded to the host CPU without any modifications. From the \bmac{} packets, the \protocolprocessor{} extracts the relevant data and forwards that data to the \blkprocessor{} for further processing. This enables efficient access to data because the \protocolprocessor{} retrieves the block and transaction data directly from the network interface in hardware, without the involvement of the host CPU.

After extracting the relevant data from the packets, the \protocolprocessor{} writes to various FIFO buffers: (1) \textbf{\blkfifo{}} every time a new block is detected, (2) \textbf{\txfifo{}} for each transaction of the block, (3) \textbf{\Endsfifo{}} for each endorsement of the transaction, (4) \textbf{\rdsetfifo{}} for each database read request of the transaction, and (5) \textbf{\wrsetfifo{}} for each database write request of the transaction.

The \textbf{\blkprocessor{}} accesses data from different buffers in parallel, and executes as many operations of validation phase as possible in parallel and pipelined fashion for high throughput. It implements block verification, transaction verification, \vscc{}, \mvcc{} and state database commit operations in the form of an integrated block-level and transaction-level pipeline. Once the \blkprocessor{} has processed the whole block, it writes to FIFO buffer \textbf{\resfifo{}}, which is read by the next module. The \textbf{\regmap{}} module writes the validation result from \resfifo{} into multiple registers in a format more suitable for software running on the host CPU. It uses AXI-Lite over PCIe as the interface with the CPU. 

Overall, the \bmac{} architecture streams block data from the network interface to \protocolprocessor{} to \blkprocessor{}, without any involvement of the host CPU. The CPU only accesses the validation result from \regmap{} to finally commit the block to disk-based ledger, as if the validation phase is offloaded to the network-attached hardware accelerator (see Figure~\ref{fig:blockchain_machine}b). The validation operations in hardware are overlapped with ledger commit in software (i.e., hardware starts processing block \textit{n}+1 while software is still committing block \textit{n}), which results in better utilization of compute resources and higher throughput.

\subsection{Protocol Processor}\label{sec:protocol_processor}
Fabric uses protobuf as the fundamental data structure for storing block and transaction data. We highlight \textbf{three reasons} why Gossip protocol (based on marshaled protobuf and gRPC/HTTP2/TCP) is not suitable as the hardware-based receiver for a network-attached hardware accelerator:
(1) Although protobufs are easily portable and heavily optimized for variable-length integer and string encoding, their decoding complexity can become problematic~\cite{parimidatacenter}. In Fabric, from our analysis, there could be up to 23 layers in the marshaled block protobuf. To retrieve a value from a protobuf embedded in a particular layer, the receiver has to recursively decode all the outer layers first, and then decode the target value itself. This results in inefficient access to relevant data in hardware.
(2) Blocks in Fabric can be as large as 100MB, which are transmitted via multiple TCP packets (see Figure~\ref{fig:validation_phase_network_stack}b). The receiver has to receive all the packets to reconstruct the block, before it could be processed. This typically requires a large buffer in receiver, which is expensive for hardware-based networking stack.
(3) We observed that a block contains quite a sizable amount of redundant data. Each block and all its transactions contain the identity of their creator, while each endorsement in a transaction contains the identity of its endorser peer, where each identity is essentially an X.509 certificate with a size of ${\sim}$860 bytes. Our analysis (see Figure~\ref{fig:protocol_results}a) shows that at least 73\% size of a block is attributed to repetitive appearance of the same identities (a typical Fabric network has just a few organizations). This makes block transmission less efficient, and the hardware has to process the same data repetitively.

\begin{figure}[t]
	\centering
	\includegraphics[width=.95\columnwidth]{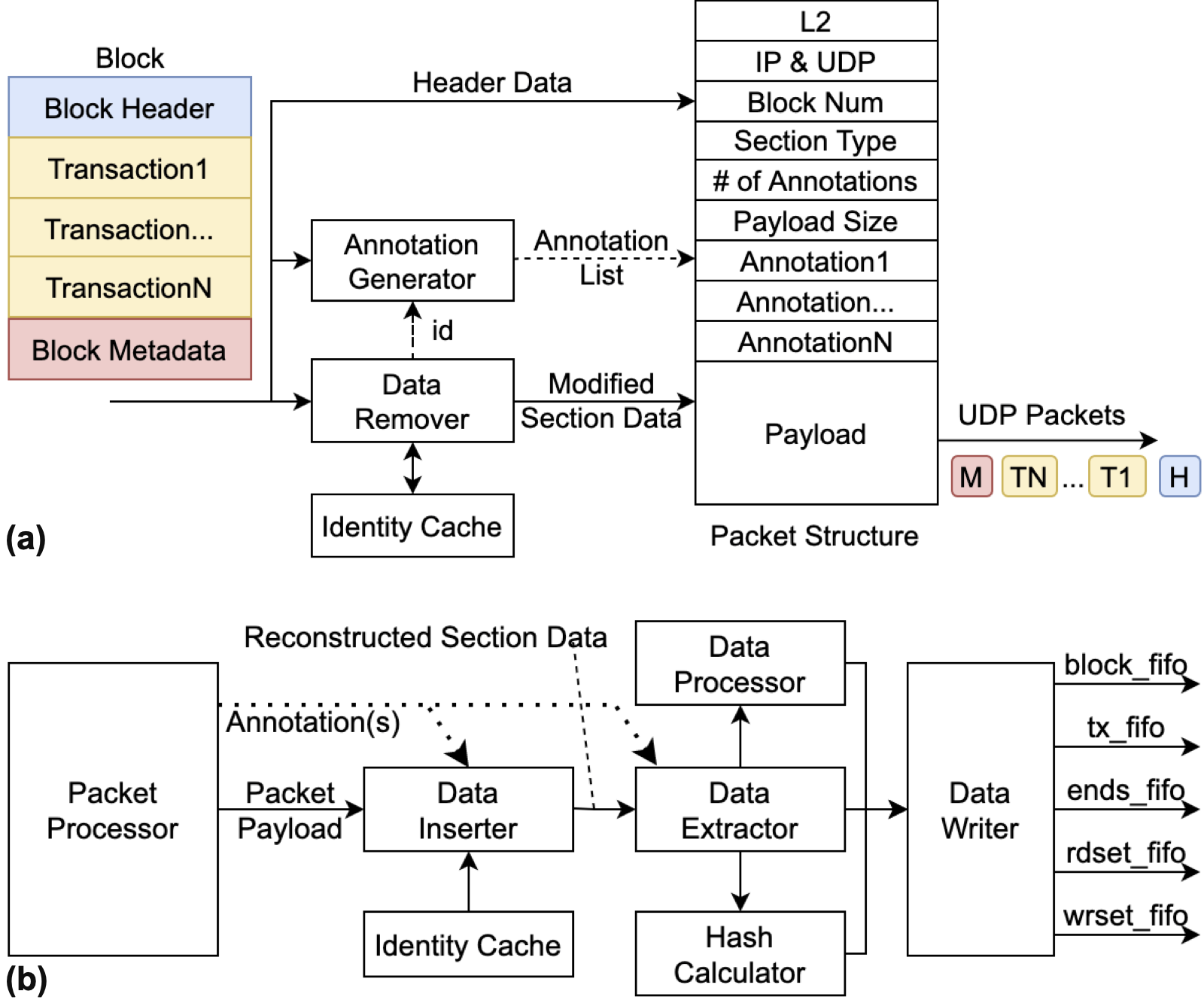}
	\caption{Blockchain Machine protocol: (a) sender flow (b) hardware-based receiver (\protocolprocessor{} module).}
	\label{fig:protocol_sender_processor}
	\vspace{-4ex}
\end{figure}

\begin{figure*}[t]
	\centering
	\includegraphics[width=2.1\columnwidth]{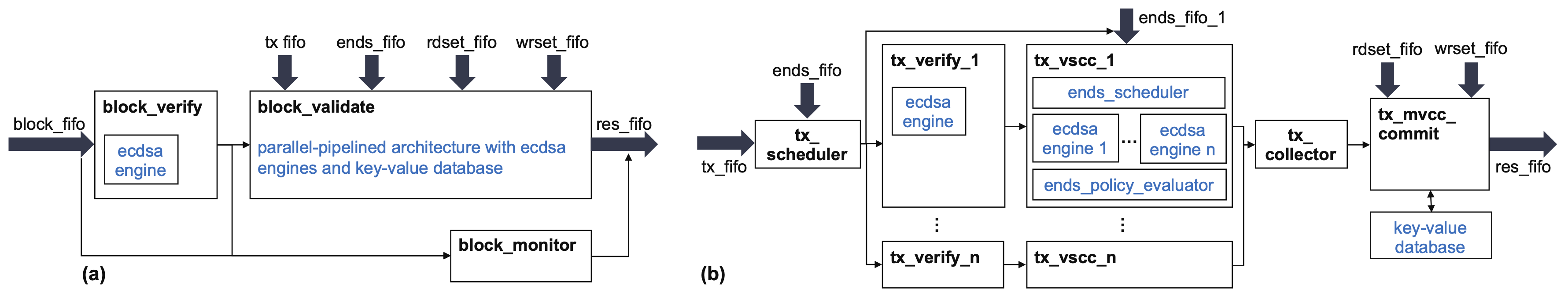}
	\caption{The \blkprocessor{} module: (a) top-level architecture (b) architecture of block\_validate stage.}
	\label{fig:block_processor}
	\vspace{-2ex}
\end{figure*}

To address the above concerns, we propose \textbf{\bmp{}}, which is a hardware-friendly communication protocol for Fabric. Fundamentally, it uses self-contained UDP packets to disseminate a block, i.e., each packet contains enough information for efficient retrieval and further processing of data from its payload, without waiting for other packets. This results in small resource footprint compared to TCP protocol where a large buffer is needed in hardware to receive the entire block before processing it. Figure~\ref{fig:protocol_sender_processor}a depicts how a block is sent through \bmp{}. Instead of treating a block as one message, it is broken down into three top-level sections: header, transactions, and metadata. The transactions section is further split into multiple sections each containing exactly one transaction. As an example, a block with 5 transactions will be broken down into 7 sections (1 header + 5 transaction sections + 1 metadata). The protocol applies two transformations to each section before sending it out in its own individual packet: (1) The original identities (i.e, certificates) are replaced with their encoded ids, which results in much smaller data being repeated in a section. (2) Annotations are inserted in the header which point to relevant data fields in a section, so that the hardware accelerator can use those annotations to know exactly where to look for when retrieving data from the packet payload.

The \textbf{\dataremover{}} module checks for presence of identities in a section, and replaces each identity with its id which is retrieved from an identity cache. If the identity is not yet present in the cache, then it is inserted with a newly assigned id. The identity cache is a map of identities (i.e., certificates) to their ids, where each id is a 16-bit integer with first 8 bits representing the organization, the next 4 bits representing one of the predefined roles in Fabric (i.e., orderer, admin, peer or client), and the last 4 bits representing the node sequence number in its organization (e.g., 0 for Org1.Peer0). This scheme results in unique ids across all the nodes of a Fabric network. Note that the number of bits used is not a limitation of our protocol; we chose 16 bits because it is enough for many Fabric networks deployed in practice. The \textbf{\annotator{}} module analyzes section data to find the offset and length of data fields required by hardware accelerator (such as block number, signatures, endorsements, etc.), and encodes them as annotations. An annotation can be either a pointer (data field offset and length) or locator (offset of removed identity and its encoded id).

Each section is sent in its own packet, which is constructed with standard L2, IP and UDP headers. The \bmp{} header is inserted as L7 header which has two parts: the fixed part contains block number, type of section in payload (i.e., header, transaction or metadata), number of annotations and the payload size, while the variable part contains the actual annotations from \annotator{} module. The payload contains the modified section data from \dataremover{} module.


Figure~\ref{fig:protocol_sender_processor}b depicts the architecture of \textbf{\protocolprocessor{}}, which is essentially a hardware-based \bmp{} receiver. For high throughput and low latency, it is designed as a packet processing system: as soon as a \bmac{} packet is received, it retrieves identities (i.e., certificates) to reconstruct original section data (\datainserter{} module), extracts/computes various data fields (\dataextractor{}, \dataprocessor{} and \hashcalculator{} modules), and writes relevant data for \blkprocessor{} (\datawriter{} module).

The \textbf{\packetprocessor{}} is used to filter out \bmac{} packets based on L3 header (i.e., type UDP with a predefined port number). For each \bmac{} packet, it parses the L7 header to retrieve all the annotations. The locator annotations and the packet payload are forwarded to the \textbf{\datainserter{}} module, which uses the encoded ids from the locator annotations to look up the original identities from a hardware-based identity cache. This cache is initialized and updated by the sender whenever a new identity is encountered.

The next module, \textbf{\dataextractor{}}, uses the pointer annotations and the reconstructed section data to extract relevant data fields. It produces three types of outputs: (1) data fields such as block number, transaction number, etc. which can be used directly and are passed to \datawriter{}, (2) data fields such as signatures, endorsements, etc. which need further processing and are passed to \dataprocessor{}, and (3) data fields that are needed for hash calculations and are passed to \hashcalculator{}. The \textbf{\dataprocessor{}} internally uses a variety of postprocessors to further extract relevant data. For example, ECDSA signatures are encoded in DER format~\cite{ITU2002} in Fabric, so the \dataprocessor{} first decodes the signature data field to find its two parts (r and s), and then coverts those parts to 256-bit values (which are expected by ECDSA verification hardware). Other postprocessors include extraction of public key from X.509 certificate, and a simplified protobuf decoder for extraction of endorsements, and read and write sets. The \textbf{\hashcalculator{}} module consists of 3 stream-based SHA-256 hash calculators. These calculators are used to compute: (1) block hash over block data (header and all transaction sections), (2) each transaction's hash over its own section, and (3) each endorsement's hash over the endorsement data in its transaction section. These hashes are used by the ECDSA verification hardware.

The last module, \textbf{\datawriter{}}, collects all the data needed by the \blkprocessor{} and writes it to various buffers. It pushes to a buffer as soon as all the data for that particular buffer is available.

\begin{figure}[t]
	\centering
	\includegraphics[width=.90\columnwidth]{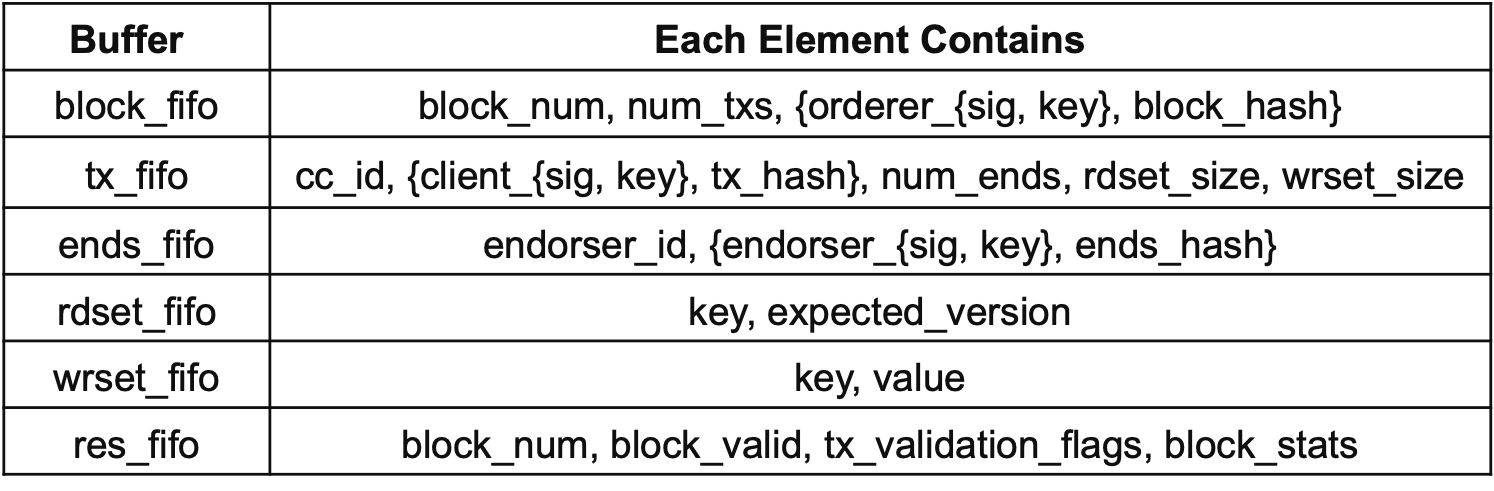}
	\caption{Various buffers used by \blkprocessor{} module (sig = signature, cc = chaincode, ends = endorsement).}
	\label{fig:buffer}
	\vspace{-2ex}
\end{figure}

\vspace{-1ex}
\subsection{Block Processor}\label{sec:block_processor}
The aim of \blkprocessor{} is to process blocks as fast as possible for high throughput.We achieve that by not only processing multiple blocks in a pipelined fashion, but also by processing multiple transactions of a block in parallel and pipelined fashion. Consequently, \blkprocessor{} is designed as an integrated block-level and transaction-level pipeline. Since Fabric uses 256-bit ECDSA scheme by default, each \textbf{\ecdsaeng{}} we use implements the same algorithm. A verification request is represented as a tuple of \textit{\{signature, key, data hash\}}, which is input to an \ecdsaeng{} for verification. Note that our architecture is not limited to ECDSA scheme; a different implementation can be plugged-in for the \ecdsaeng{}.

\textbf{Block-level Pipeline.} Figure~\ref{fig:block_processor}a depicts the top-level architecture of \blkprocessor{} and the contents of its buffers are described in Figure~\ref{fig:buffer}. To avoid large buffers, instead of using long ids from Fabric to differentiate between transactions and their data, we exploit the fact that \protocolprocessor{} writes to these buffers in the same order as it receives the blocks and their transactions. That is, we use num\_txs from \blkfifo{} to read that many transactions from \txfifo{}, essentially reading only the transactions of a single block. Likewise, we use rdset\_size/wrset\_size from \txfifo{} to pop that many database read/write requests from \rdsetfifo{}/\wrsetfifo{}.

The \blkprocessor{} is organized as a 2-stage block-level pipeline, where the first stage \textbf{\blkverify{}} verifies the block by reading its verification request from \blkfifo{}. One \ecdsaeng{} is dedicated for this stage to ensure that blocks are verified as soon as they arrive. The \textbf{\blkvalidate{}} stage validates all the transactions of a block, commits its valid transactions, and writes to \resfifo{} after the entire block has been processed. The statistics in \resfifo{} are collected by the \textbf{\blkmonitor{}} module, which monitors the FIFO buffers and interfaces of the stages to keep track of the time spent in various operations. Overall, the block-level pipeline processes two blocks at a time in pipelined fashion.

\textbf{Transaction-level pipeline.} Figure~\ref{fig:block_processor}b depicts the internal architecture of \blkvalidate{} stage, which consists of a 3-stage transaction-level pipeline. For high throughput, the transaction-level pipeline is based on two principles: (1) Since transaction processing is dominated by signature verifications and \ecdsaeng{} takes much longer (typically the case with cryptographic operations) than other modules, it is extremely important to execute verification requests only when necessary. We use various early abort conditions along the pipeline to skip a transaction as soon as it becomes invalid (e.g., endorsements are discarded if transaction failed verification or endorsement policy was invalidated). (2) Process as many transactions in parallel as possible, so we distribute many \ecdsaeng{} instances across two pipeline stages to create an efficient parallel-pipelined architecture.

The first stage \textbf{\txverify{}} verifies the transaction using its verification request from \txfifo{}. One \ecdsaeng{} is dedicated to this stage to ensure that transactions of a block are verified as soon as the block has been verified. The \txverify{} implements an internal mechanism to skip the transaction when its already invalid (i.e., when the block has already been marked as invalid by \blkverify{}). The \txverify{} produces the transaction valid/invalid status, along with its cc\_id, num\_ends, rdset\_size and wrset\_size (which were read from \txfifo{}; see Figure~\ref{fig:buffer}) for the next stage.

The second stage \textbf{\txvscc{}} implements verification of endorsements of a transaction, and checks whether those endorsements satisfy endorsement policy of the associated chaincode. If the transaction was marked invalid by \txverify{}, then all of its endorsements are discarded, and the transaction is passed on as invalid. We use multiple yet configurable number of \ecdsaeng{} instances in \txvscc{} to process multiple endorsements of a transaction in parallel. The \textbf{\Endsscheduler{}} issues a new endorsement as soon as a free \ecdsaeng{} instance is available. This ensures that all the instances are kept busy as long as there are unprocessed endorsements, and hence are not under-utilized. The \Endsscheduler{} also collects the output of an \ecdsaeng{}; it forwards current transaction's cc\_id along with the valid/invalid status of the endorsement and its associated endorser\_id (endorser peer's encoded id from Section~\ref{sec:protocol_processor}) to the \Endspolicyeval{} module.

As mentioned earlier, we should verify only as many endorsements as needed to satisfy an endorsement policy. For example, a transaction with ``2-outof-3 orgs'' policy will have three endorsements (one from each of the three organizations), and as soon as the first two endorsements are found to be valid, the transaction should be marked valid while skipping the third endorsement. We design such a fast evaluation scheme with combinational circuits for endorsement policies and short-circuit evaluation.

The \textbf{\Endspolicyeval{}} consists of a register file, where each register represents an organization and each register bit represents one of the predefined roles (i.e., orderer, admin, peer or client). Consider a Fabric network with four organizations, then there will be four registers each 4-bit wide. Since endorser ids are encoded ids (with both organization and role of an endorser peer), \Endspolicyeval{} uses the endorser id to write the endorsement verification result to the corresponding register bit. Assuming reg1 = Org1, reg2 = Org2 and bit2 = peer role, then the verification result of endorsement from Org1.Peer will be written to reg1 bit2. This enables us to use a combinational circuit for parallel evaluation of an endorsement policy. Assuming peer roles in ``2-outof-3 orgs'' endorsement policy, it can be represented as ``(Org1.Peer \& Org2.Peer) | (Org1.Peer \& Org3.Peer) | (Org2.Peer \& Org3.Peer)''. We implement the boolean expression as a combinational circuit where the inputs are bits from the register file. Using the earlier example, for ``(Org1.Peer \& Org2.Peer)'' expression, reg1 bit2 and reg2 bit2 will be connected to the input of an AND gate. The entire endorsement policy can be implemented using three 2-input AND gates and one 3-input OR gate. Since each chaincode has its own endorsement policy, each endorsement policy is implemented as a separate combinational circuit. The cc\_id of the current transaction is used to select the circuit representing its endorsement policy, which then becomes the output of \Endspolicyeval{}. 

The short-circuit evaluation is implemented in the \Endsscheduler{} module. Since endorsement policies are boolean expressions, short-circuit evaluation can be applied to check for validity (i.e., stop as soon as the policy is satisfied) or invalidity (i.e., stop as soon as the policy is not satisfied). We argue that the majority of transactions are meant to be valid in real-life applications, and thus it is more efficient to apply short-circuit evaluation for validity of an endorsement policy. When \txvscc{} starts a new transaction, the register file in \Endspolicyeval{} is cleared, which means that the default status of an endorsement policy is invalid (i.e., not satisfied). The \Endsscheduler{} checks the output of \Endspolicyeval{} before issuing any new endorsements. If the output is valid, then it means the endorsement policy has been satisfied and the remaining endorsements can be discarded. If all the endorsements have been processed, and the output of \Endspolicyeval{} is still invalid, then it means the endorsement policy cannot be satisfied and the transaction is marked as invalid.

To further enhance throughput, we use a configurable number of \txverify{} and \txvscc{} instances to process multiple transactions in parallel. The \textbf{\txscheduler{}} issues a new transaction of the current block as soon as a free \txverify{} instance is available, i.e., read transaction data from \txfifo{} and forward it to the selected \txverify{} instance, and issue transaction endorsements to the connected \txvscc{} instance. Since multiple transactions are processed in parallel, it is possible that they may finish out-of-order because some transactions may have less number of endorsements than others or their endorsement policies are satisfied earlier. However, the \mvcc{} operation expects the transactions in the original order because it has to sequentially check for read-write conflicts. Therefore, we introduce the \textbf{\txcollector{}} module, which collects the output of all the \txvscc{} instances in-order, i.e., it will collect the output of a \txvscc{} instance only if that output corresponds to transaction tx1, then apply the same check for tx2, and so on until all the transactions of the block are collected.

The third stage \textbf{\txmvcc{}} implements the \mvcc{} operation and commits to state database. Since cryptographic operations take much longer than other operations, we combined \mvcc{} and commit operations in one stage for a more balanced pipeline. We use an in-hardware key-value store as the state database, which supports read and write operations. The read operation takes a key as input, while the write operation takes a key and value pair as input. The value field contains the actual value and a version number, to enable storage of versioned data~\cite{Androulaki2018}. The database implements an internal locking mechanism to disallow reading of a key if it is currently being written. Although the in-hardware database could be small for real-life applications, see Section~\ref{sec:discussion} for possibilities to use a larger database.


The input to \txmvcc{} is block number, sequence number of transaction within the block, rdset\_size and wrset\_size. It reads from \rdsetfifo{}, and issues database read request for each of those entries. A transaction is marked as valid only if the version from the database response of each key matches with the corresponding expected version from the \rdsetfifo{}. If the transaction is valid, then its \wrsetfifo{} entries are committed to the database by issuing a write request for each entry consisting of key and value (which is a combination of original value from \wrsetfifo{} and its version created from block number and transaction sequence number~\cite{Androulaki2018}). The commit operation is skipped for transactions marked invalid by \mvcc{}, while both the \mvcc{} and commit operations are skipped for transactions already marked invalid by \txvscc{} stage.

\textbf{Adaptability.} The \blkprocessor{} is a parallel-pipelined architecture, capable of achieving high throughput because multiple transactions in parallel stream through the pipeline to be committed to the database, and to produce the validation result of the entire block. The \blkprocessor{} can be adapted using a number of configurable options:
(1) If an application uses a large block size, then we can use more parallel instances of \txverify{} and \txvscc{} to achieve high throughput.
(2) If the endorsement policy of a chaincode requires verification of 3 endorsements in most of the transactions, then we can configure 3 \ecdsaeng{} instances in \txvscc{} to best handle the common case.
(3) The \Endspolicyeval{} is generated automatically (see Section~\ref{sec:implementation_details}) using the chaincode endorsement policies, so multiple smart contracts can be supported.

\vspace{-1ex}
\subsection{Hardware/Software Interface}\label{sec:hwsw_interface}
The \textbf{\regmap{}} module is the primary interface between the \bmac{} architecture and any software running on the host CPU. Internally, it consists of multiple registers that are accessible at predefined addresses. These registers store the validation result of a block, which consists of block number, block valid/invalid status, number of transactions in the block, transactions' valid/invalid flags, and block statistics. The \regmap{} also uses a mechanism to block writing of new data to the registers until the previous data has been read by the CPU, which ensures that validation result of a block is not overwritten before it is accessed by the CPU.

Figure~\ref{fig:blockchain_machine}b shows how the validation phase is executed in \bmac{} peer. When \bmac{} peer receives a block from the orderer sent through our protocol, the \protocolprocessor{} extracts block data and its transactions for validation in hardware (right-hand side). The same block is also received by the Fabric peer software running on the host CPU, either through the Gossip protocol (since this is just like any other peer in the Fabric network) or from our \protocolprocessor{} in hardware (which can forward UDP packets containing block data to the CPU as well). The software still executes some parts of the block and transaction verification operations which are not suitable for hardware accelerator, as shown on the left-hand side. However, after that, the software skips all the remaining operations until the ledger commit operation. At this point, the software reads validation result of the block from hardware, and combines it with the original block, and then commits the updated block to ledger just like the software-only peer.

\vspace{-1ex}
\subsection{Implementation Details}\label{sec:implementation_details}
We implemented \bmac{} architecture in OpenNIC~\cite{Xilinx2020OpenNIC}, which is an open-source NIC platform for Xilinx accelerator cards. The OpenNIC platform provides Ethernet interface for network connectivity, DMA and AXI-Lite interfaces through PCIe for host CPU connectivity, and a user box for implementation of hardware accelerators. The \bmac{} architecture is implemented as the user box, where the \protocolprocessor{} is connected to the Ethernet interface and the \regmap{} is connected to the AXI-Lite interface. The implementation of BMac architecture features a multi-language combination of RTL, HLS, and P4~\cite{Bosshart2014}, used according to their strengths. For example, \packetprocessor{} is implemented using P4, simpler modules such as \datawriter{} are implemented using HLS, and \blkprocessor{} is implemented in RTL. The in-hardware database is implemented using BRAM/URAM, and hence is entirely on FPGA.

A YAML based configuration file is used to define both static and configurable parameters of \bmac{}. For example, it contains identity information (certificates, roles, etc.) of various nodes of the Fabric network, and chaincode endorsement policies. We use a script which parses this configuration file to generate encoded ids for peers (as explained in Section~\ref{sec:protocol_processor}), and automatically generates the \Endspolicyeval{} module using those ids and endorsement policies (as explained in Section~\ref{sec:block_processor}). Note that the \Endspolicyeval{} can be regenerated to update the \bmac{} architecture when organizations or endorsement policies change.

The protocol we proposed is implemented in Go language as a library to hide the low-level details of encoding block data and synchronizing identity cache. The protocol uses \bmac{} configuration file mentioned earlier to generate encoded ids. The primary function is Send(), which takes a block as input in the same format (i.e., marshaled protobuf) as used by Fabric software. The orderer is modified to call Send() right before a new block is sent through Gossip protocol. Therefore, the same orderer can send blocks to both software-only and \bmac{} peers. Only the lead orderer in multi-node Raft ordering service sends the block through our protocol.

We implemented an API in Go language to hide the low-level interactions with \bmac{} hardware. Other than the standard open/close functions to access the FPGA card, the primary function is GetBlockData(). This function returns when the \regmap{} has the validation result, and converts that result into a format compatible with Fabric peer software. The peer software is modified to skip all the operations offloaded to the \bmac{} hardware, and calls GetBlockData() right before the ledger commit operation. Our modifications are guarded by a flag, so the modified software can be used either as software-only or \bmac{} peer by just changing that flag.

\section{Evaluation}\label{sec:experiments}

\begin{figure}[t]
	\centering
	\includegraphics[width=0.95\columnwidth]{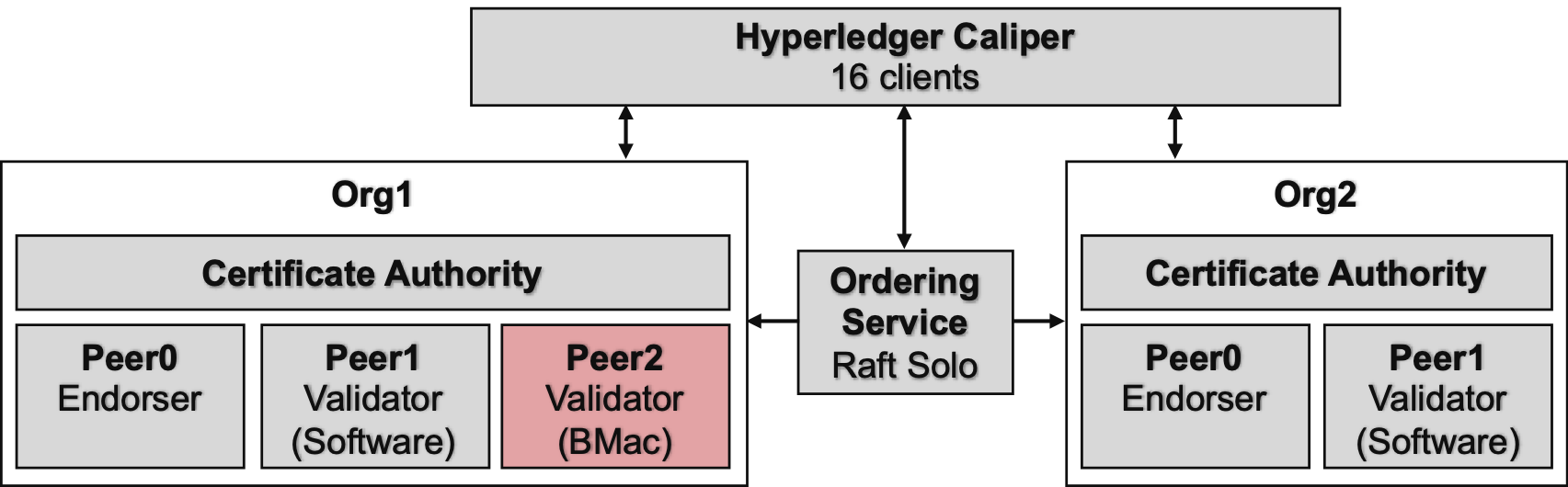}
	\caption{Experimental Setup.}
	\label{fig:experimental_setup}
\vspace{-4ex}
\end{figure}

\subsection{Experimental Setup}
We use Hyperledger Caliper~\cite{HyperledgerCaliper}, the standard benchmarking tool for Fabric, to evaluate the performance of \bmac{} peer. We instrumented Fabric peer code to log timestamps through the validation phase, which are then gathered by Caliper to report statistics at the block-level to better evaluate the validation phase~\cite{Gorenflo2019FastFabric,Javaid2019}. For \bmac{} peer, block statistics read from the hardware are reported. We used Fabric v1.4.5 with LevelDB in all our experiments.

Figure~\ref{fig:experimental_setup} shows the typical setup used in our experiments. We created a Fabric network with a single channel and two organizations, each having a certificate authority, one endorser and one software-only validator (\textbf{\swvalidator{}}) peer. In Org1, we added \bmac{} peer, which acts just like any other validator peer, illustrating the compatibility goal from Section~\ref{sec:introduction}. Raft based ordering service is used with a single orderer. We created various virtual machines (VMs), where each VM is equipped with Intel Xeon 4414 @ 2.2GHz vCPUs and 2GB RAM per vCPU. Caliper is run on a 16 vCPU VM with 16 clients (for higher transaction send rates), while the ordering service and all the certificate authorities are run on a VM with 8 vCPUs. All the software-only peers are run in their individual VMs with varying number of vCPUs (and the same number of \vscc{} threads to validate multiple transactions in parallel). However, \bmac{} peer is always run on a 4 vCPU VM because its software only commits blocks to disk-based ledger. All the VMs are connected through a 1Gbps network.


The \bmac{} architecture is configured to support blocks containing up to 256 transactions, and database with 8192 entries. Multiple architectures are generated with chaincode endorsement policies of the benchmarks and varying number of \textbf{\txvalidators{}} (parallel instances of \txverify{} + \txvscc{} where each \txvscc{} has 2 instances of \ecdsaeng{} unless stated otherwise). All \bmac{} architectures are implemented with a target frequency of 250MHz. We also implemented a high-level simulator for \bmac{} architecture, and use it only to measure performance of architectures beyond 16 \txvalidators{}. The performance reported by our simulator is always within 1\% of actual measurements from the hardware. To ensure that our \bmac{} implementation did not alter the behavior of validation phase, in each experiment, we compared block and transactions' valid/invalid flags, and commit hash (based on block data) between the software-only and \bmac{} peers. We did not find any mismatches in our experiments.

\subsection{Metrics and Benchmarks}
Since our goal is to evaluate performance of validation phase, we use \textbf{commit throughput}, which is defined as the rate at which transactions are committed by the peer, and \textbf{block validation latency}, which is the total time taken by the peer to validate and commit the entire block. For direct comparison between hardware and software implementations, we exclude ledger commit operation (similar to~\cite{Gorenflo2019FastFabric}) when computing these metrics because it is always executed on CPU.

We used \textbf{smallbank} and \textbf{digital rights management (drm)} applications from Caliper benchmarks~\cite{HyperledgerCaliperBenchmarks}, which have been used as representative benchmarks in literature~\cite{Gorenflo2019FastFabric,Javaid2019,Zhu2020}. The smallbank application implements typical functions of a banking application, while the drm application implements typical functions of managing digital assets. The endorsement policy is set to ``2-outof-2 orgs'' unless stated otherwise. Caliper clients create random transactions, and a total of 150,000 transactions (30,000 repeated 5 times) are used to compute average metrics of an experiment.



\subsection{Results}
\begin{figure}[t]
\vspace{-4ex}
	\centering
	\includegraphics[width=\columnwidth]{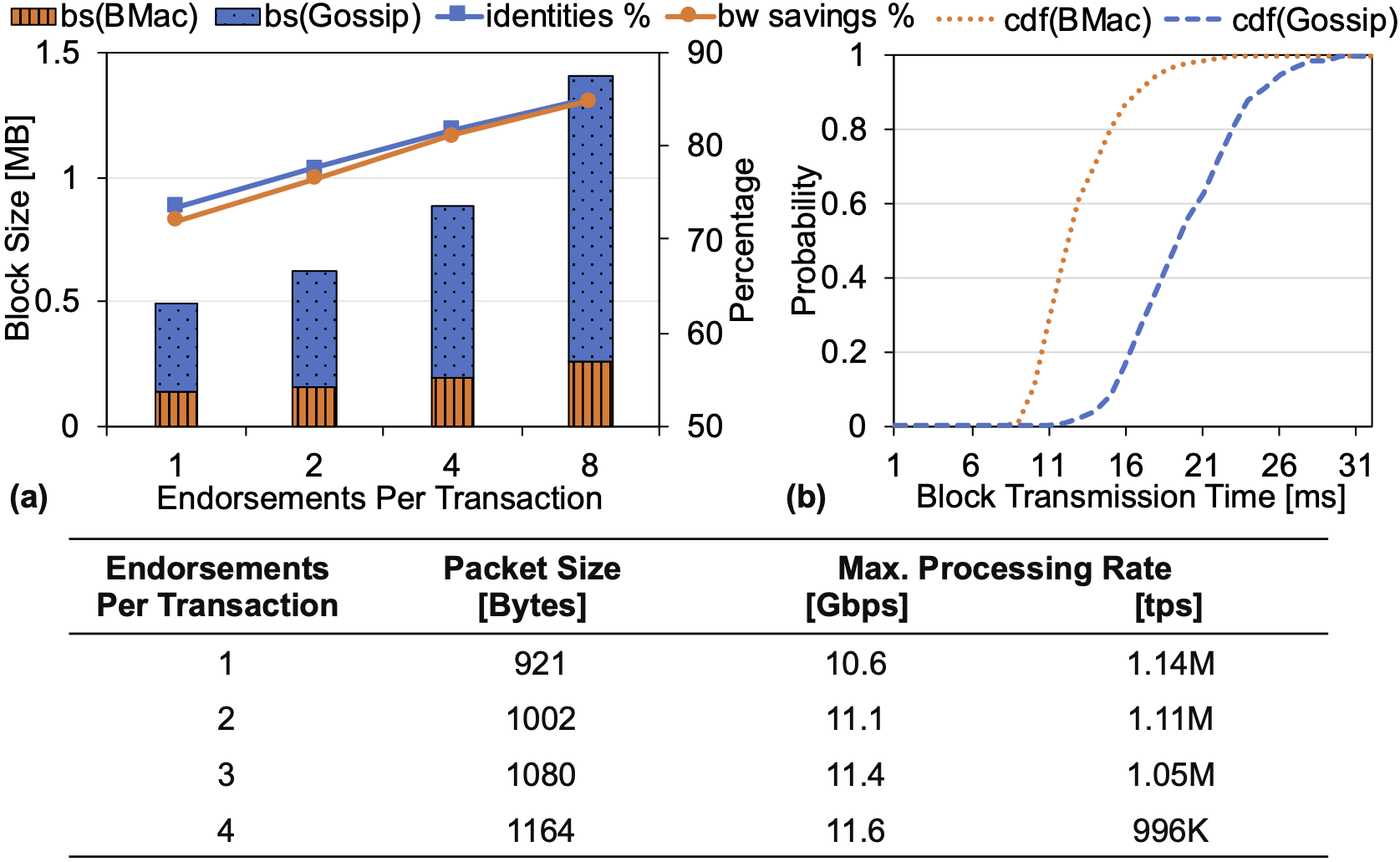}
	\caption{Blockchain Machine protocol performance.}
	\label{fig:protocol_results}
\end{figure}

\textbf{Protocol Performance.} We first report the performance of \bmac{} protocol, by collecting data of at least 500 blocks each with 150 transactions (\textasciitilde10mins of experiment running time). Figure~\ref{fig:protocol_results}a shows the network bandwidth savings of \bmac{} vs. Gossip protocol. With more endorsements per transaction, the block size in Gossip increases significantly because of more and more identity certificates (which constitute at least 73\% of a block). In contrast, the corresponding block size in \bmac{} protocol is 3.4$\times$ -- 5.3$\times$ smaller, resulting in bandwidth savings of up to 85\%. The table below the figure reports performance of the \protocolprocessor{} module, which is capable of processing incoming data up to a rate of 11Gbps, which translates to at least 996,000 tps. This shows that both the \bmac{} protocol sender and hardware-based receiver (\protocolprocessor{}) are very fast and can sustain high throughput. Note that more endorsements means larger transaction packet, and hence less number of transactions being sent in the same time period, lowering throughput in tps. Figure~\ref{fig:protocol_results}b reports CDF of end-to-end block transmission time. The 95th percentile of \bmac{} protocol is 18ms compared to 26ms of Gossip protocol, which is a reduction of 30\% in latency.


%

\begin{figure}[t]
	\centering
	\includegraphics[width=0.9\columnwidth]{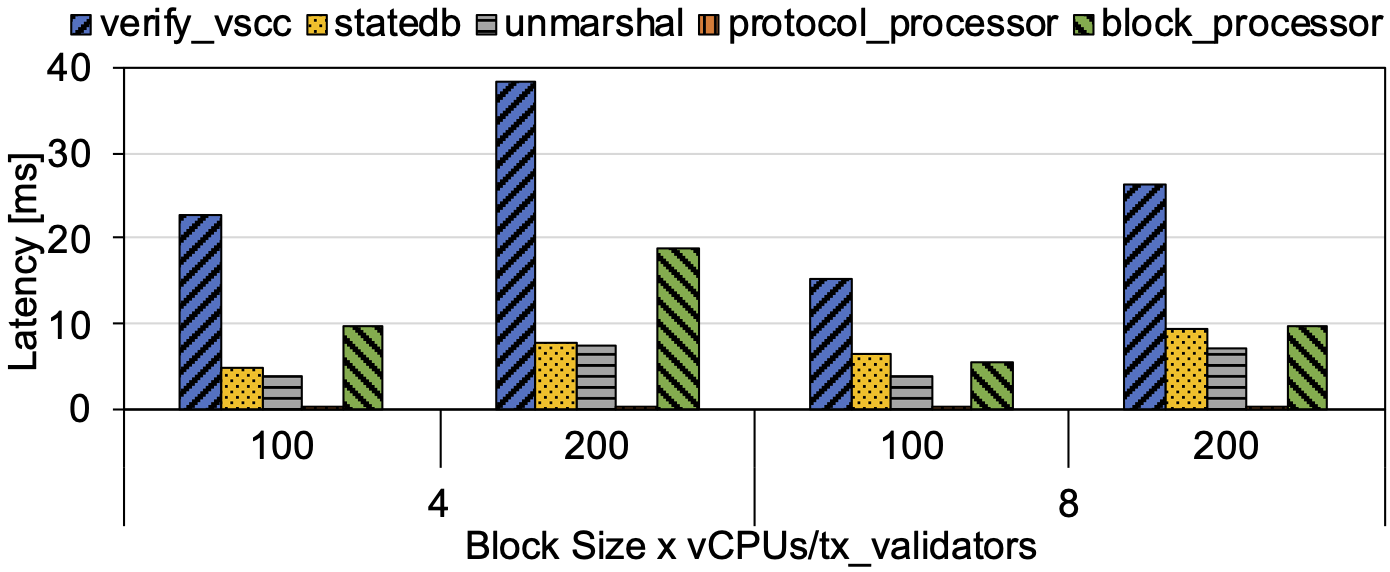}
	\caption{Breakdown of block validation for \swvalidator{} and \bmac{} peers.}
	\label{fig:speedup}
	\vspace{-2ex}
\end{figure}

\textbf{High Throughput.} First, we present breakdown of block validation latency in Figure~\ref{fig:speedup} to highlight the benefits of accelerating various bottleneck operations. We changed the block size (number of transactions), and vCPUs and \txvalidators{} for \swvalidator{} and \bmac{} peers respectively. The verify\_\vscc{} and statedb operations from \swvalidator{} peer are equivalent to \blkprocessor{}, while unmarshal operation (retrieval of block and transactions data) is equivalent to \protocolprocessor{} in \bmac{} peer. For block size 200 and 8 vCPUs/\txvalidators{}, latency to parse and retrieve block data is improved by ${\sim}$40$\times$ to less than 0.2ms by \protocolprocessor{}, while block validation is improved by ${\sim}$3.7$\times$ (from 35.9ms to 9.7ms) by \blkprocessor. This clearly illustrates the efficiency of removing large certificates from a block and using self-contained UDP packets in our protocol, as well as parallel-pipelined processing of blocks and transactions in \blkprocessor{}. Since unmarshaling accounts for ${\sim}$17\% of validation latency, while verify\_\vscc{} and statedb operations account for more than 80\%, these improvements translate to an overall 4.4$\times$ speedup in block validation for \bmac{} peer vs. \swvalidator{} peer. These improvement trends are similar across block sizes and number of vCPUs/\txvalidators{}, so we only report overall performance for the rest of this section.

\begin{figure}[b]
	\vspace{-2ex}
	\centering
	\includegraphics[width=1.0\columnwidth]{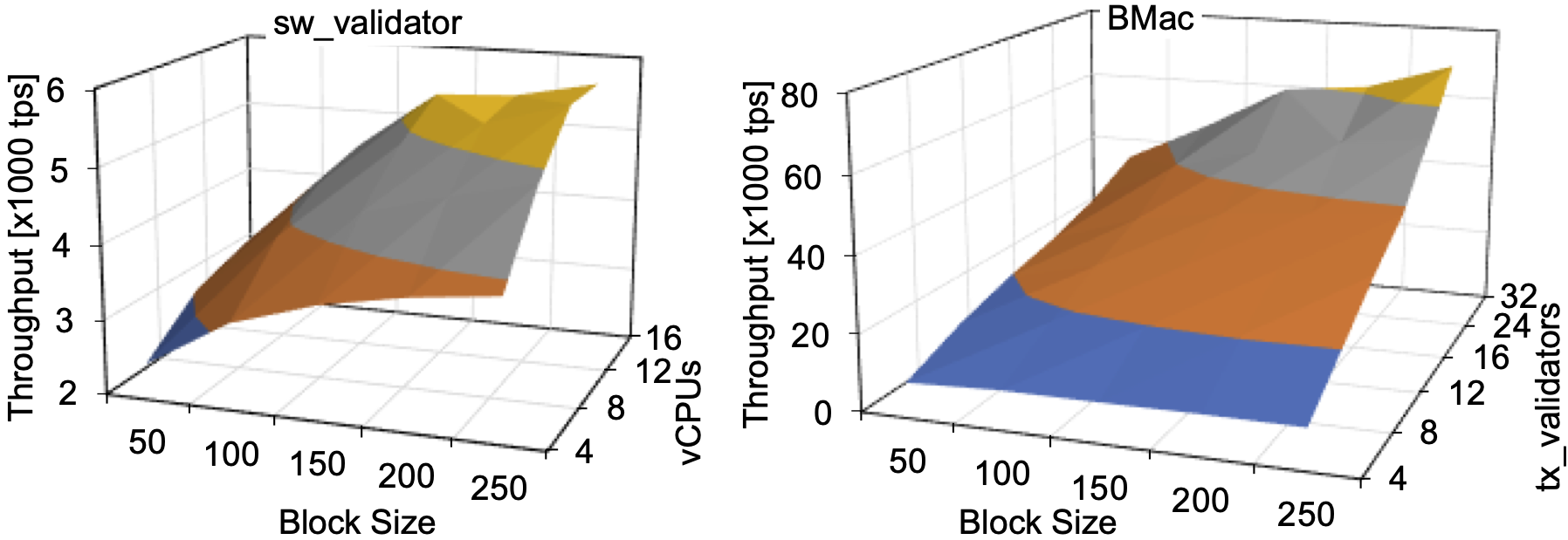}
	\caption{Smallbank results with different block sizes, vCPUs and \txvalidators{}.}
	\label{fig:perf_bs_cpu}
\end{figure}

\begin{figure*}[t]
	\centering
	\includegraphics[width=2.05\columnwidth]{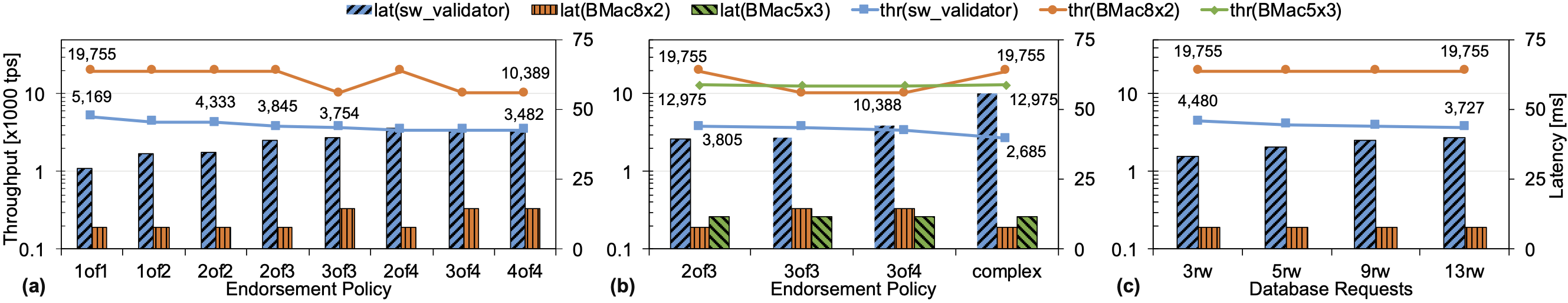}
	\caption{Smallbank results with different endorsement policies and database requests.}
	\label{fig:perf_policies_database}
	\vspace{-2ex}
\end{figure*}

For overall performance, we changed block size, vCPUs for \swvalidator{} peer and \txvalidators{} for \bmac{} peer. The results are reported in Figure~\ref{fig:perf_bs_cpu}. For both peers, the throughput increases with larger block sizes because the fixed cost of processing a block (e.g., protobuf unmarshaling for \swvalidator{} or filling up of transaction-level pipeline for \bmac{}) is better amortized. For block size 250, the throughput of \bmac{} peer increases by 3.6$\times$ (10,700 tps to 38,400 tps) with 4 to 16 \txvalidators{}. When compared to ideal speedup of 4$\times$, this improvement illustrates that \bmac{} efficiently uses the provided compute resources. In contrast, the throughput of \swvalidator{} peer increased by a mere 1.5$\times$ (3,900 tps to 5,600 tps). This is because \swvalidator{} peer only uses parallel threads for \vscc{} operation, and \mvcc{} and commit operations are executed sequentially without any pipelining, limiting the maximum throughput achievable. Interestingly, \bmac{} peer with only 4 \txvalidators{} achieved 2$\times$ the throughput of \swvalidator{} peer with even 16 vCPUs (10,700 tps vs. 5,600 tps).

Overall, \bmac{} peer could deliver throughput of up to 68,900 tps with block latency of 3.63ms (not shown in the figure). Considering 16 vCPUs as a common multi-core server configuration, this is ${\sim}$12$\times$ improvement over 5,600 tps of \swvalidator{}. Note that the throughput of \swvalidator{} barely improved after 16 vCPUs, thus is not reported. For perspective, the best reported throughput in literature is 20,000 tps from FastPeer~\cite{Gorenflo2019FastFabric} (14,000 tps from its stable implementation~\cite{Gorenflo2020FastFabricImplementation}), and an estimated throughput of 16,800 tps from SmartFabric~\cite{Thakkar2021} (5,600 tps from \swvalidator{} scaled by 3$\times$ reported improvement). Furthermore, transaction validation latency in \bmac{} peer is ${\sim}$0.7ms, which is about the same as the best reported in StreamChain~\cite{Kuhring2020}, however,~\cite{Kuhring2020} achieved a throughput of only 3,100 tps. To further highlight the potential of \bmac{}, we used our simulator with even higher number of \txvalidators{}. The \bmac{} peer could deliver throughput of ${\sim}$100,000 tps for block size 250 with 50 \txvalidators{}, and ${\sim}$150,000 tps for block size 500 with 80 \txvalidators{}. Hence, one can choose larger FPGAs for implementation to achieve the highest performance.

\textbf{Adaptability.} In these experiments, we change the endorsement policies and hence the number of endorsements on transactions, which translates to varying crypotgraphic workload in \vscc{} operation. The results are reported in Figure~\ref{fig:perf_policies_database}a where eight different endorsement policies (2of2 means ``2-outof-2 orgs'' policy) are used with 8 vCPUs/\txvalidators{} and block size of 150. Compared to the setup in Figure~\ref{fig:experimental_setup}, we added Org3 and Org4, each with one endorser peer. The throughput of \swvalidator{} peer decreases almost linearly with the number of endorsements; throughput with 3of3 policy is 13.5\% less than that of 2of2 policy, which is 16\% less than that of 1of1 policy. When comparing 18.3ms \vscc{} latency (not shown in figure) of 1of1 policy to 23.2ms for 2of2 policy and 28ms for 3of3 policy, it turns out that evaluation of one more endorsement takes about 5ms, but the fixed cost of policy evaluation is quite high (${\sim}$13ms). The throughput with 2of3 and 3of3 policies is about the same (3,800 tps), which comes as a surprise because 2 endorsements should have been enough for 2of3 policy. It turns out that Fabric always verifies all the endorsements of a transaction, irrespective of the policy. In contrast, the short-circuit evaluation in \bmac{} architecture ensures only necessary endorsements are evaluated; the throughput of 19,800 tps with 2of3 policy is much higher than 10,400 tps with 3of3 policy. 

The critical stage in \bmac{} architecture is \txvscc{} because an \ecdsaeng{} takes much longer (${\sim}$360us per verification~\cite{MercurySystems2021}) than the rest of the operations (tens of us). Therefore, \bmac{} architecture is more sensitive to the number of verifications performed: for 3of3 policy with 2 \ecdsaeng{} instances in \txvscc{}, the first 2 endorsements are verified in parallel and then another iteration is needed to verify the third endorsement, almost doubling \vscc{} latency. However, we can leverage the adaptability of our architecture to customize it for the applications. Figure~\ref{fig:perf_policies_database}b compares two \bmac{} architectures with about the same number of total \ecdsaeng{} instances, but organized differently: 8x2 and 5x3 (5 \txvalidators{} each with 3 instances of \ecdsaeng{} in \txvscc{}). The 8x2 architecture outperforms by 52\% for 2of3 policy, while the 5x3 architecture outperforms by 25\% for 3of3 and 3of4 policies. Therefore, one should use 8x2 and 5x3 architectures for applications using 2ofN and 3ofN policies, respectively. We also evaluated a complex endorsement policy ``(Org1 \& Org2) \textbar{} (Org1 \& Org4) \textbar{} (Org2 \& Org3) \textbar{} (Org2 \& Org4) \textbar{} (Org3 \& Org4)'' which is almost but not exactly like 2of4 policy. From Figure~\ref{fig:perf_policies_database}b, the \swvalidator{} peer throughput decreased significantly to ${\sim}$2,700 tps because Fabric implementation evaluates all sub-expressions of a policy sequentially. In contrast, \bmac{} architecture uses combinational circuits which inherently evaluate sub-expressions of a policy in parallel. Hence, \bmac{} throughput is almost the same as 2of4 policy (19,800 tps).

\textbf{Other Results.} We modified smallbank application to include the functionality of split payment to \textit{n} accounts, resulting in variable number of database reads and writes (rw). Figure~\ref{fig:perf_policies_database}c reports the results with 8 vCPUs/\txvalidators{} and block size of 150. The \bmac{} peer throughput remains the same (19,800 tps); although the latency of \txmvcc{} stage increased with more database accesses but it is still hidden by the latency of \txvscc{} stage. In contrast, the \swvalidator{} peer throughput decreased by a total of 16\%.

\begin{figure}[t]
	\centering
	\includegraphics[width=\columnwidth]{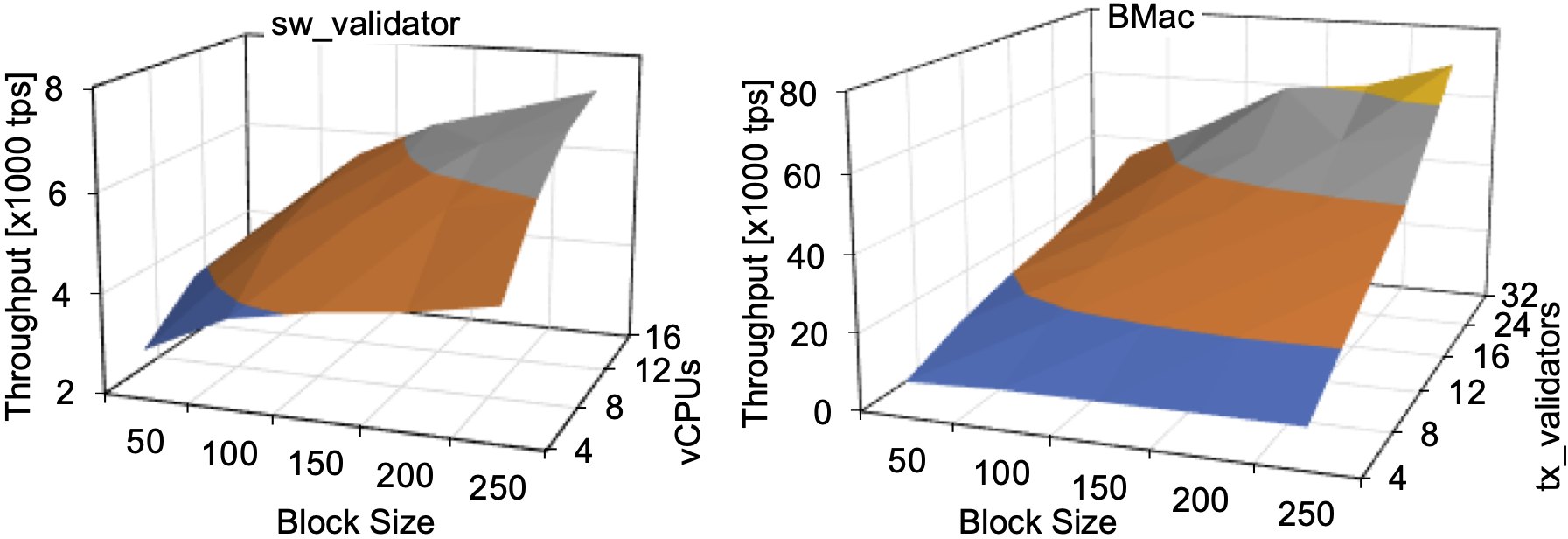}
	\caption{Drm benchmark results.}
	\label{fig:perf_drm}
\vspace{-3ex}
\end{figure}

Figure~\ref{fig:perf_drm} shows the results for drm application. The trends are similar to those of smallbank, so only a subset of the results is reported. The \swvalidator{} achieved better throughput than smallbank because drm application has less accesses to database, which means faster \mvcc{} and commit operations. The throughput of \bmac{} peer is very similar to smallbank because its dominated by \vscc{} latency, which is the same in both applications due to the same endorsement policy.

\textbf{Hardware Resource Utilization.} Table~\ref{tab:hardware_resources} reports the utilization of different \bmac{} architectures on Xilinx Alveo U250 accelerator card~\cite{Xilinx2020Alveo}. Even the largest \bmac{} architecture 16x2 uses less than half of the FPGA resources. The utilization of GT, BUFG, MMCM and PCIe resources is 83.3\%, 2.2\%, 6.3\% and 25\% respectively, and is about the same for all architectures. Since \txvscc{} is the critical stage in \bmac{} architecture, the extra FPGA resources available can be used to improve latency of an \ecdsaeng{} and/or add more \ecdsaeng{} instances for better throughput.

\begin{table}[b!]
	\vspace{-2ex}
	\centering
	\begin{tabular}{cccccc}
		\hline
		\textbf{Resource} & \textbf{4x2} & \textbf{5x3} & \textbf{8x2} & \textbf{12x2} & \textbf{16x2} \\ \hline
		LUT / LUTRAM & 20.9\% & 25.4\% & 28.5\% & 35.8\% & 43.3\% \\
		FF & 6.9\% & 7.3\% & 8.0\% & 9.1\% & 10.3\% \\
		BRAM / URAM & 13.1\% & 13.1\% & 13.1\% & 13.1\% & 13.1\% \\
		\hline
	\end{tabular}
	\caption{Hardware utilization of \bmac{} architectures.}
	\label{tab:hardware_resources}
\end{table}
\section{Discussion}\label{sec:discussion}
In this paper, we used an in-hardware key-value database, whose size is limited by the available FPGA resources. For real-life applications, a much larger database may be needed. One option is to use in-hardware database for small amount of actively accessed data, while keeping a persistent database on the host CPU. Several existing techniques~\cite{Sakakibara2018,Xu2016,Memcached} can be used to create such a setup. Since \txvscc{} is the critical stage in \bmac{} architecture, increased database access latencies over PCIe in \txmvcc{} stage (when a larger database is kept on the host) could still be hidden by \ecdsaeng{} latency from \txvscc{} stage. This will be addressed in a future work as an enhancement to the platform proposed here.


In our protocol, we did not propose or implement a retransmission scheme for lost packets. The bandwidth requirement of our protocol is much lower than the typical link bandwidth available within a datacenter. Therefore, existing schemes such as Go-Back-N can be used as it has been used in RDMA over Ethernet~\cite{guo2016rdma}. If Fabric nodes are deployed across datacenters (typically not the case), then we can leverage a TCP based proxy for WAN communication, and transmit block data from proxy to \bmac{} peer using our protocol (since both of them will be within the same datacenter). These will be addressed in a future work as an enhancement to the protocol proposed here. If each section of a block cannot fit into the standard Ethernet frame, we can opt for larger Ethernet frames~\cite{EC2jumbo}. Like Gossip, our protocol can also be used by the lead peer to send blocks to other peers in its own organization.


Although we integrated \bmac{} hardware with Fabric v1.4 LTS, our proposals are applicable to Fabric v2.2 LTS because the validation phase is fundamentally the same between these versions. When newer chaincodes are installed in a Fabric network, the \bmac{} peer must be restarted for reprogramming it with the updated hardware architecture. One solution is to use partial reconfiguration to reprogram only the endorsement policy evaluator module without restarting the peer.

\section{Related Work}\label{sec:related_work}
In this section, we report the most relevant literature on improving Fabric peer performance. Readers are referred to~\cite{Anh2018,Thakkar2018,Javaid2019,Chung2019,Zhu2020} for benchmarking/tuning studies and~\cite{Androulaki2018} for consensus mechanism.

Many recent works have proposed optimizations to improve peer performance, which include parallel validation of transactions~\cite{Thakkar2018,Gorenflo2019FastFabric,Thakkar2021}, caching unmarshaled blocks~\cite{Gorenflo2019FastFabric}, pipelined execution of validation/commit operations~\cite{Gorenflo2019FastFabric,Kuhring2020,Thakkar2021}, partial validation/commit of blocks~\cite{Thakkar2021}, separating endorsement and validation phases~\cite{Gorenflo2019FastFabric}, parallel/asynchronous state database and ledger commits~\cite{Javaid2019,Gorenflo2019FastFabric}, and considering transaction conflicts during block creation/validation~\cite{Sharma2019,Ruan2020,Lee2019,Gorenflo2019Xox}. All these optimizations are software-based, and hence are limited by the computational power of underlying general-purpose resources such as CPUs in multi-core servers. In contrast, we proposed a hardware/software co-designed peer, leveraging both CPUs and FPGA-based accelerator to deliver performance beyond the software-only wall. Note that some of these optimizations can be applied to our architecture to further improve throughput, e.g., validation/commit of partial block from~\cite{Thakkar2021} can be implemented as a transaction filter in \txscheduler{} module.

Our protocol reduces network bandwidth (not only for hardware but also for software implementations) by removing redundant identity certificates, which has not been explored before. Compared to~\cite{berendea2020fair}, which optimized Gossip protocol to reduce its long tail propagation and improve its bandwidth utilization, our protocol focuses on faster transmission and decoding of blocks in hardware. The authors of~\cite{parimidatacenter,diamantopoulos2020phryctoria} proposed hardware-based protobuf decoders, while~\cite{Zhang2017} proposed a protobuf-like messaging framework with associated hardware accelerators. However, just using hardware-based protobuf decoder is not enough due to the complex/layered structure of block in Fabric. The authors of~\cite{Sakakibara2018} proposed an FPGA-based key-value store for a blockchain-based digital assets application. Unlike our work,~\cite{Sakakibara2018} does not accelerate Fabric's validation phase which is much more than just database accesses.

\section{Conclusion}\label{sec:conclusion}
We proposed Blockchain Machine which is a network-attached hardware accelerator with its own protocol for Hyperledger Fabric's validation phase. The Blockchain Machine enables efficient transfer and access of block data, and efficient verification and validation of transactions. It can be adapted to applications as different chaincode endorsement policies can be compiled into hardware and the transaction-level pipeline is configurable according to cryptographic workload of the endorsement policies. Our experiments illustrated that Blockchain Machine peer is capable of achieving high commit throughput, low block validation latency, and is adaptable and scalable. Possible future works include scaling the in-hardware database and upgrading to Fabric v2.2 LTS.


\section{Acknowledgements}
We would like to thank Mercury Systems for providing us the ECDSA verification IP.

\bibliographystyle{ACM-Reference-Format}
\bibliography{library_haris,library_ji}

\end{document}